\documentclass[journal,twoside,final]{IEEEtran}

\usepackage{amsmath,amssymb,amsfonts,amsthm}
\usepackage{balance}
\usepackage{bm}
\usepackage{cite}
\usepackage{enumitem}
\usepackage[T1]{fontenc}
\usepackage{graphicx}
\usepackage{mathrsfs}
\usepackage{mathtools}
\usepackage{multirow}
\usepackage{orcidlink}
\usepackage{siunitx}
\usepackage{soul}               
\usepackage{threeparttable}
\usepackage[color=green!40]{todonotes}
\soulregister\cite7
\soulregister\footnote7
\soulregister\eqref7
\soulregister\ref7
\usepackage{xcolor}
\usepackage{url}
\usepackage{hyperref}
\newlength{\myvspace}
\newtheorem{definition}{Definition}
\newtheorem{lemma}{Lemma}
\newtheorem{theorem}{Theorem}

\DeclareMathOperator*{\argmin}{arg\,min}

\DeclareMathOperator*{\fp}{fp}

\newcommand{\td}{\triangledown}

\graphicspath{{./}{../figures/}}

\newcommand{\myauthor}{Kyeong Soo
  Kim$^{\orcidlink{0000-0002-4123-2647}}$,~\IEEEmembership{Senior~Member,~IEEE}}%
\newcommand{\mytitle}{Direct Search Algorithm for Clock Skew Compensation Immune
  to Floating-Point Precision Loss}%

\begin{document}

\title{\LARGE \mytitle}

\author{%
  \myauthor%
  \thanks{%
    This work was supported in part by the Postgraduate Research Scholarships
    (under Grant FOSA2412040 and PGRS1912001) of Xi'an Jiaotong-Liverpool
    University.

    K. S. Kim is with the Department of Communications and Networking, School of
    Advanced Technology, Xi'an Jiaotong-Liverpool University, Suzhou 215123,
    P. R. China (e-mail: Kyeongsoo.Kim@xjtlu.edu.cn).%
  }%
}%


\maketitle

\begin{abstract}
  We have been investigating clock skew compensation immune to floating-point
  precision loss by taking into account the discrete nature of clocks in digital
  communication systems; extending Bresenham's line drawing algorithm, we
  constructed an incremental error algorithm using only integer
  addition/subtraction and comparison. Still, bounding the initial value of the
  clock remains a challenge, which determines the initial condition of the
  algorithm and thereby its number of iterations. In this letter, we propose a
  new incremental error algorithm for clock skew compensation, called direct
  search, which no longer relies on the bounds on the initial value of the
  clock. The numerical examples demonstrate that the proposed algorithm can
  significantly reduce the number of iterations in comparison to the prior work
  while eliminating the effect of floating-point precision loss on clock skew
  compensation.
\end{abstract}


\section{Introduction}
\label{sec:introduction}
\IEEEPARstart{N}{oting} that the performance of typical clock skew compensation
algorithms based on floating-point operations does not match with the prediction
from theories or simulation experiments due to the use of 32-bit
single-precision floating-point format on resource-constrained platforms like
battery-powered, low-cost wireless sensor network (WSN) nodes~\cite{Kim:20-1},
we have been investigating clock skew compensation immune to floating-point
precision loss. Specifically, we proposed an incremental error algorithm based
on the extension of Bresenham's line drawing
algorithm~\cite{bresenham65:_algor}, which takes into account the discrete
nature of clocks in digital communication systems and thereby eliminates the
effect of limited floating-point precision on clock skew
compensation~\cite{Kim:22-1}. We further complemented the algorithm by providing
practical as well as theoretical bounds on the initial value of a
skew-compensated clock by floating-point division based on systematic analyses
of the errors of floating-point operations~\cite{Kang:23}.

The numerical examples in~\cite{Kang:23}, however, show that both theoretical
and practical bounds are looser than the approximate bounds of~\cite{Kim:22-1}
and thereby increase the number of iterations of the algorithm. In this letter,
we propose a new incremental error algorithm for clock skew compensation, called
direct search, which does not rely on the bounds on the initial value of the
clock.

\section{Review of Clock Skew Compensation Based on the Extended Bresenham's
  Algorithm}
\label{sec:csc-ext-bresenham}
As in~\cite{Kim:22-1,Kang:23}, we consider only the clock skew compensation at a
sensor node, whose hardware clock $T$ is described as follows with respect to
the reference clock $t$ of the head node:
\begin{equation}
  \label{eq:hw_clock_model}
  T(t) = \left(1+\epsilon\right)t
\end{equation}
where $\epsilon{\in}\mathbb{R}$ is the clock skew. Compensating for the clock
skew from the hardware clock $T$ in \eqref{eq:hw_clock_model}, we can obtain the
logical clock $\hat{t}$ of the sensor node---i.e., the estimation of the
reference clock $t$ given the hardware clock $T$---as follows: For
$t_{i}{<}t{\leq}t_{i+1}~(i{=}0,1,{\ldots})$,
\begin{equation}
  \label{eq:logical_clock_model}
  \hat{t}\Big(T(t)\Big) = \hat{t}\Big(T(t_{i})\Big)
  + \dfrac{T(t)-T(t_{i})}{1 + \hat{\epsilon}_{i}},
\end{equation}
where $t_{i}$ is the reference time for the $i$th synchronization between the
head and the sensor nodes and $\hat{\epsilon}_{i}$ is the estimated clock skew
from the $i$th synchronization. The major issue is the floating-point division
required for the calculation of the second term of rhs in
\eqref{eq:logical_clock_model}, i.e.,
$\frac{T(t){-}T(t_{i})}{1{+}\hat{\epsilon}_{i}}$, which is the skew-compensated
increment of the hardware clock since the $i$th synchronization.

Because clocks in digital communication systems are basically discrete counters
and timestamps exchanged among nodes are their values, we proposed an
incremental error algorithm to obtain the second term of rhs in
\eqref{eq:logical_clock_model} using only integer addition/subtraction and
comparison by extending the Bresenham's line drawing algorithm
in~\cite{Kim:22-1,Kang:23}: If the inverse of a clock frequency ratio (i.e.,
$\frac{1}{1{+}\epsilon_{i}}$) is estimated and described as $\frac{D}{A}$ with
two positive integers $D$ and $A$, the clock skew can be compensated for based
on Theorem~1:
\begin{theorem}[Clock Skew Compensation Based on the Extended Bresenham's
  Algorithm~\cite{Kim:22-1,Kang:23}]
  \label{thm:csc_with_optimal_bounds}
  Given a hardware clock $i$, we can obtain its skew-compensated clock $j$ as
  follows:
	
  \smallskip%
  \noindent
  \textit{Case 1.} $\frac{D}{A}{<}1$: The skew-compensated clock $j$ satisfies
  \begin{equation}
    \label{eq:csc-scc_bounds}
    \left\lfloor{i\frac{D}{A}}\right\rfloor \leq j \leq \left\lceil{i\frac{D}{A}}\right\rceil.
  \end{equation}
  Unless $i\frac{D}{A}$ is an integer, there are two values satisfying
  \eqref{eq:csc-scc_bounds}. Because we cannot know the exact value of
  $i\frac{D}{A}$ due to limited floating-point precision, however, we extend
  \eqref{eq:csc-scc_bounds} to include the effect of the precision loss: For
  floating-point numbers with a base 2 and a precision $p$\footnote{$p$ is
    precision in bits; for example, $p{=}24$ for the 32-bit single-precision
    floating-point format (i.e., binary32) defined in IEEE
    754-2008~\cite{IEEE:754-2008}.},
  \begin{equation}
    \label{eq:csc-scc_optimal_bounds}
    \left\lfloor{\frac{1{-}u{+}2u^{2}}{(1{+}u)^{2}(1{+}2u)}t}\right\rfloor \leq j \leq
    \left\lceil{\frac{(1+2u)^{3}(1{+}u{-}2u^{2})}{(1{+}u)^{2}}t}\right\rceil,
  \end{equation}
  where $t{=}i\frac{D}{A}$ and $u{=}2^{-p}$.
	
  Let $k,{\ldots},k{+}l$ be the candidate values of $j$ satisfying
  \eqref{eq:csc-scc_optimal_bounds}. We determine $j$ by applying the extended
  Bresenham's algorithm of~\cite{Kim:22-1} with $\Delta{a}$ and $\Delta{b}$ set
  to $A$ and $D$ from the point $(i{-}l,k)$ and on; $j$ is determined by the $y$
  coordinate of the valid point whose $x$ coordinate is $i$.
	
  \smallskip%
  \noindent
  \textit{Case 2.} $\frac{D}{A}{>}1$: In this case, we can decompose the
  skew-compensated clock $j$ into two components as follows:
  \begin{equation}
    \label{eq:decomposition}
    j = i\frac{D}{A} = i + i\frac{D - A}{A}.
  \end{equation}
  Now that $\frac{D{-}A}{A}{<}1$, we can apply the same procedure of Case~1 to
  the second component in \eqref{eq:decomposition} by setting $\Delta{a}$ and
  $\Delta{b}$ of the extended Bresenham's algorithm to $A$ and $D{-}A$,
  respectively. Let $\bar{j}$ be the result from the procedure. The
  skew-compensated clock $j$ is given by $i{+}\bar{j}$ as per
  \eqref{eq:decomposition}.
\end{theorem}

\begin{figure}[!tb]
  \begin{center}
    \includegraphics[angle=-90,width=\linewidth]{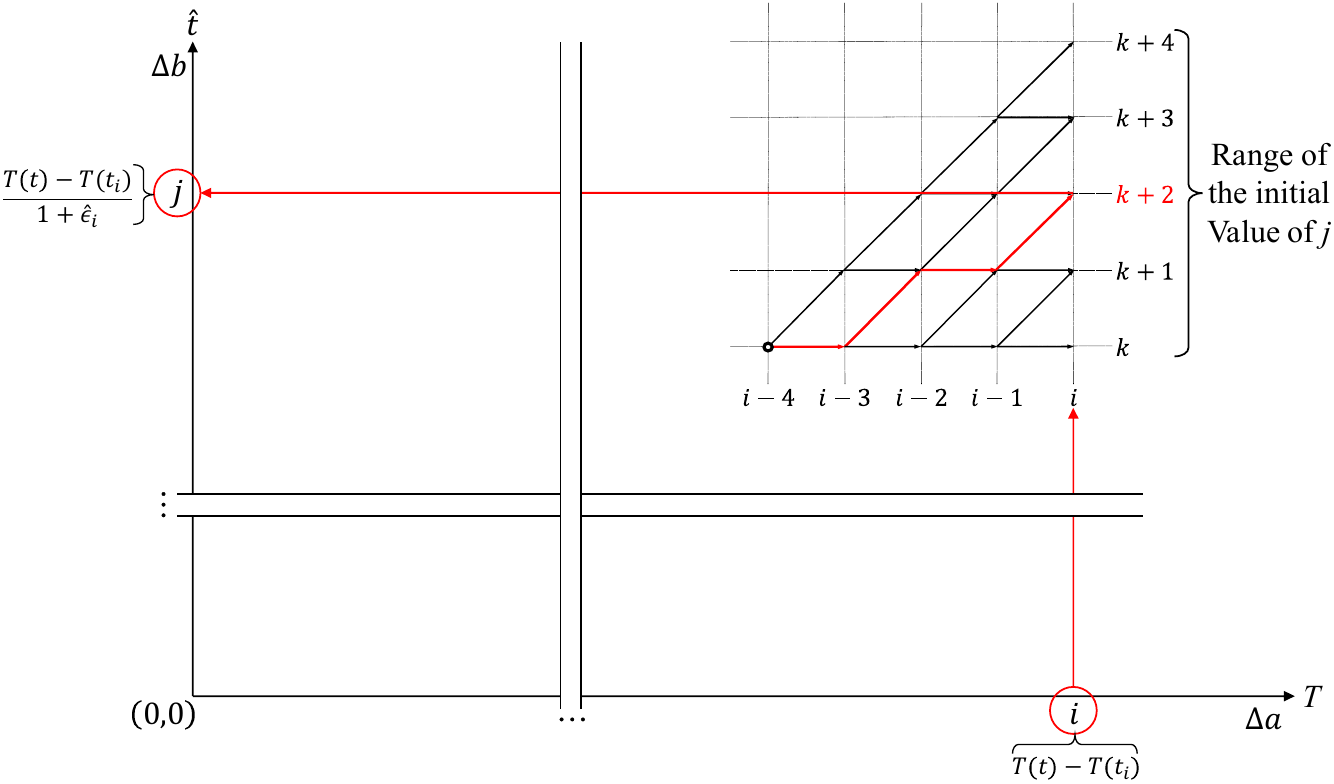}
  \end{center}
  \caption{Clock skew compensation based on the extended Bresenham's
    algorithm~\cite{Kim:22-1,Kang:23} for the case of $\frac{D}{A}{<}1$.}
  \label{fig:csc-ext-bresenham}
\end{figure}

Note that practical bounds loosening the theoretical bounds in
\eqref{eq:csc-scc_optimal_bounds} are also provided for practical implementation
at resource-constrained sensor nodes with limited floating-point precision
in~\cite{Kang:23}. As illustrated in Fig.~\ref{fig:csc-ext-bresenham}, the range
of the initial value of $j$, which are bounded in
\eqref{eq:csc-scc_optimal_bounds}, determines the number of iterations of the
algorithm (i.e., the difference between the upper and lower bounds).
The numerical examples presented in~\cite{Kang:23} demonstrate that, though
being based on systematic analyses of the errors of floating-point operations,
the theoretical and practical bounds are unnecessarily looser than the
approximate bounds of~\cite{Kim:22-1} and, therefore, increase the number of
iterations, which is the motivation for the current work on a new incremental
error algorithm not relying on the bounds on the initial value of the clock.

\section{Direct Search Algorithm}
\label{sec:direct-search}
First, we formally define the skew-compensated clock of a given hardware clock
as follows:
\begin{definition}
  \label{def:scc}
  Let $\mathbb{N}_{0}$ be a set of non-negative integers. Given a hardware clock
  $i{\in}\mathbb{N}_{0}$, we define its skew-compensated clock $j$ as follows:
  \begin{equation}
    \label{eq:scc-1}
    j \triangleq \argmin_{k \in \mathbb{N}_{0}}\left|k-i\frac{D}{A}\right|.
  \end{equation}
\end{definition}

Note that we cannot directly obtain $j$ by converting $i\frac{D}{A}$ to an
integer (e.g., rounding it half up to ${\lfloor}i\frac{D}{A}{+}0.5{\rfloor}$);
we can obtain only ${\fp}(i\frac{D}{A})$ from the floating-point operations of
$i\frac{D}{A}$ on a real hardware platform due to the effect of floating-point
precision loss, where $\fp(\cdot)$ denotes the result of the floating-point
operations on the given platform.

From Definition~\ref{def:scc}, we can derive the following lemma:
\begin{lemma}
  \label{lem:scc-cond}
  $j$ is the skew-compensated clock of a hardware clock $i$ iff $j$ satisfies
  the following condition:
  \begin{equation}
    \label{eq:scc-2}
    \left|jA-iD\right| \leq \left|kA-iD\right|
    ~~~ \forall k \in \mathbb{N}_{0}.
  \end{equation}
\end{lemma}
\begin{IEEEproof}
  If $j$ is the skew-compensated clock of the hardware clock $i$, $j$ satisfies
  the following by Definition~\ref{def:scc}:
  \begin{equation}
    \label{eq:scc-3}
    \left|j-i\frac{D}{A}\right| \leq \left|k-i\frac{D}{A}\right| ~~~ \forall k \in \mathbb{N}_{0}.
  \end{equation}
  Multiplying both sides of \eqref{eq:scc-3} by $A$, we obtain \eqref{eq:scc-2}.
  
  If $j$ satisfies \eqref{eq:scc-2}, we obtain \eqref{eq:scc-3} by dividing both
  sides of \eqref{eq:scc-2} by $A$, which means that $j$ is the skew-compensated
  clock of the hardware clock $i$ according to Definition~\ref{def:scc}.
\end{IEEEproof}
\vspace{\myvspace}%

The importance of Lemma~\ref{lem:scc-cond} is that, unlike \eqref{eq:scc-1},
there is no floating-point division in \eqref{eq:scc-2}, which could be a
building block for the construction of an incremental error algorithm like the
Bresenham's algorithm~\cite{bresenham65:_algor}. Theorem~\ref{thm:csc-ds} shows
how to obtain the skew-compensated clock $j$ given the hardware clock $i$ based
on Definition~\ref{def:scc} and Lemma~\ref{lem:scc-cond} through such an
incremental error algorithm.
\begin{theorem}[Clock Skew Compensation Based on Direct Search]
  \label{thm:csc-ds}
  Given the hardware clock $i$, we can obtain its skew-compensated clock $j$ as
  follows:
  
  \smallskip%
  \noindent
  \textbf{Initialization}:
  \begin{align}
    k & = \left\lfloor \fp\!\left(i\frac{D}{A}\right) + 0.5 \right\rfloor. \label{eq:csc-ds-k} \\
    \td & = kA - iD. \label{eq:csc-ds-td}
  \end{align}
  
  \smallskip%
  \noindent
  \textbf{Recursive steps}:
  \begin{enumerate}[label={Case$\,$\arabic*},align=left]
  \item $\td{=}0$:
    \begin{itemize}
    \item $j{\leftarrow}k$ and stop.
    \end{itemize}
  \item $\td{>}0$:
    \begin{enumerate}[label*=.\arabic*,align=left]
    \item $\td{-}A{=}0$:
      \begin{itemize}
      \item $j{\leftarrow}k{-}1$ and stop.
      \end{itemize}
    \item $\td{-}A{>}0$:
      \begin{itemize}
      \item $k{\leftarrow}k{-}1$ and $\td{\leftarrow}\td{-}A$.
      \item Go to Case$\,$2.
      \end{itemize}
    \item $\td{-}A{<}0$:
      \begin{itemize}
      \item If $|\td{-}A|{<}|\td|$, $j{\leftarrow}k{-}1$; otherwise,
        $j{\leftarrow}k$.
      \item Stop.
      \end{itemize}
    \end{enumerate}
  \item $\td{<}0$:
    \begin{enumerate}[label*=.\arabic*,align=left]
    \item $\td{+}A{=}0$:
      \begin{itemize}
      \item $j{\leftarrow}k{+}1$ and stop.
      \end{itemize}
    \item $\td{+}A{>}0$:
      \begin{itemize}
      \item If $|\td{+}A|{<}|\td|$, $j{\leftarrow}k{+}1$; otherwise,
        $j{\leftarrow}k$.
      \item Stop.
      \end{itemize}
    \item $\td{+}A{<}0$:
      \begin{itemize}
      \item $k{\leftarrow}k{+}1$ and $\td{\leftarrow}\td{+}A$.
      \item Go to Case$\,$3.
      \end{itemize}
    \end{enumerate}
  \end{enumerate}
\end{theorem}
\begin{IEEEproof}
  See Appendix~\ref{sec:proof-theorem-1}.
\end{IEEEproof}
\vspace{\myvspace}%

Unlike the clock skew compensation based on the extended Bresenham's algorithm,
the direct search algorithm of Theorem~\ref{thm:csc-ds} makes no distinction
between the cases of $\frac{D}{A}{<}1$ and $\frac{D}{A}{>}1$ and handles them in
an integrated way. To avoid overflow during the multiplication of $kA$ or $iD$
with 32-bit integers, $\td$ in \eqref{eq:csc-ds-td} can be alternatively
initialized as follows:
\begin{equation*}
  \td = kA - iD = (k-i)A + i(A-D),
\end{equation*}
where both $(k{-}i)$ and $(A{-}D)$ are close to zero because
$\frac{D}{A}{\approx}1$. 

\begin{figure}[!tb]
  \begin{center}
    \includegraphics[angle=-90,width=.95\linewidth]{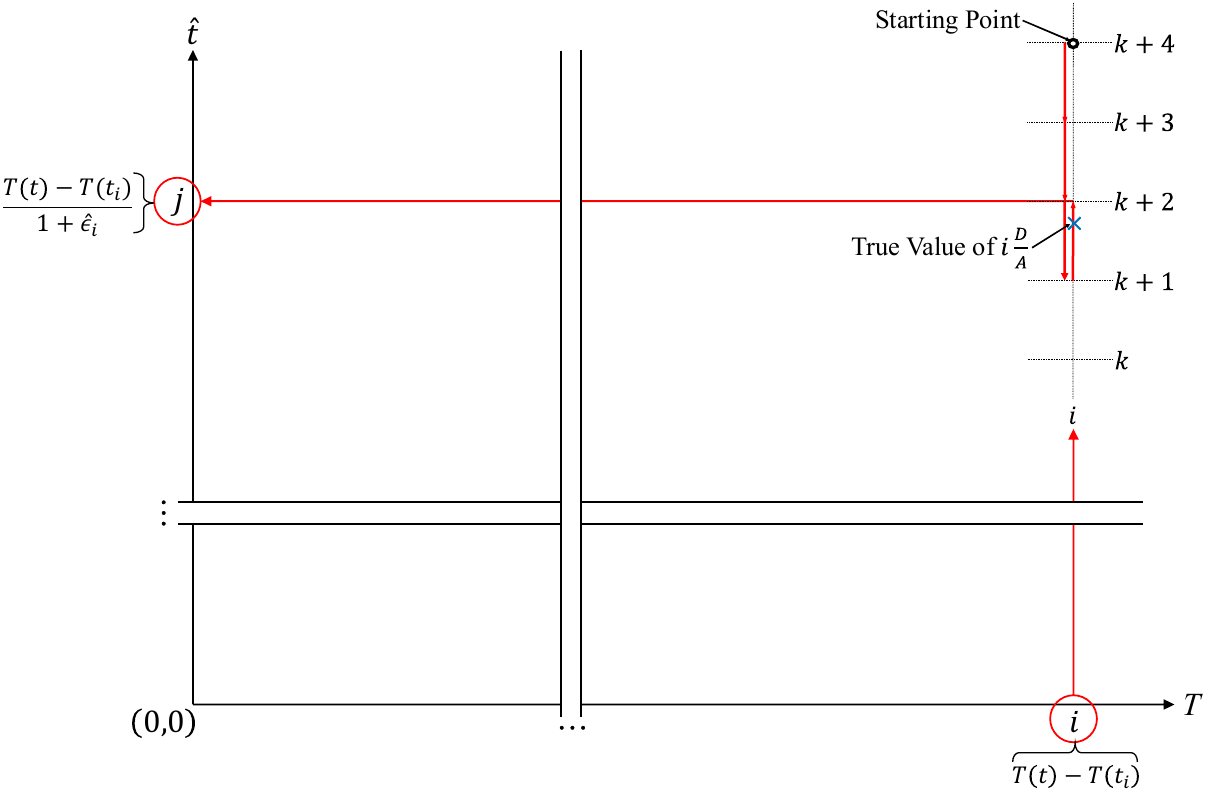}
  \end{center}
  \caption{Clock skew compensation based on the direct search algorithm.}
  \label{fig:csc-ds}
\end{figure}

\begin{figure}[!tb]
  \centering%
  \includegraphics[width=.8\linewidth]{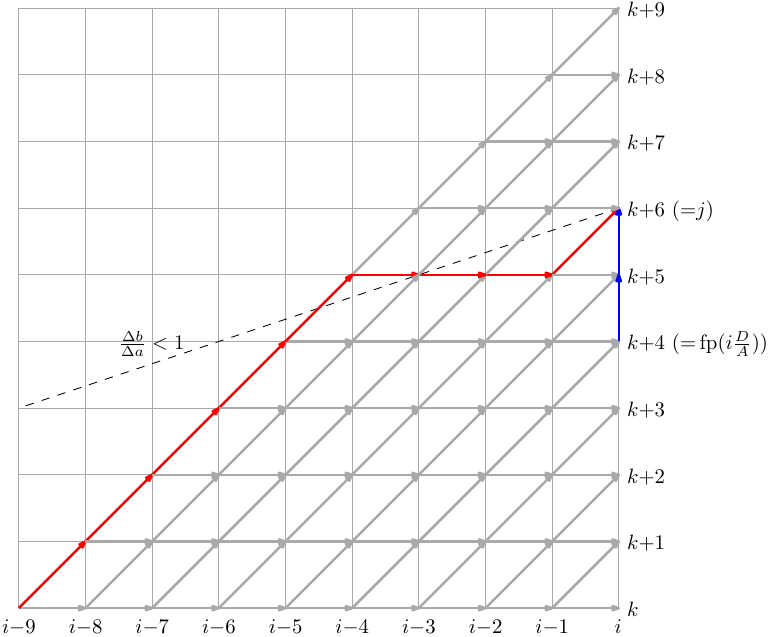}
  \caption{Comparison of the recursive steps between the direct search algorithm
    (in blue) and the extended Bresenham's algorithm~\cite{Kim:22-1,Kang:23} (in
    red).}
  \label{fig:csc-example}
\end{figure}

Fig.~\ref{fig:csc-ds} illustrates the clock skew compensation based on the
direct search algorithm, while Fig.~\ref{fig:csc-example} compares the recursive
steps between the direct search algorithm and the extended Bresenham's
algorithm.

\subsection{Numerical Examples}
\label{sec:numerical-examples}
We compare the results of clock skew compensation algorithms under the same
condition as~\cite{Kim:22-1,Kang:23}, which are summarized in
Table~\ref{tab:csc_results}.
\begin{table*}[!tb]
  \centering
  \begin{threeparttable}
    \caption{Comparison of clock skew compensation algorithms.}
    \label{tab:csc_results} {%
      \renewcommand{\arraystretch}{1.3} \begin{tabular}{c|r|r|r|r|r|r|r} \hline
        \multirow{2}{*}{Algorithm} & \multicolumn{1}{c|}{\multirow{2}{*}{$i$}} & \multicolumn{3}{c|}{Compensation error\tnote{*}} & \multicolumn{3}{c}{$\#$ of iterations} \\
        \cline{3-8}
                                   & & \multicolumn{1}{c|}{Min.} & \multicolumn{1}{c|}{Max.} & \multicolumn{1}{c|}{Avg.} & \multicolumn{1}{c|}{Min.} & \multicolumn{1}{c|}{Max.} & \multicolumn{1}{c}{Avg.} \\
        \hline \multirow{4}{*}{Single precision\tnote{\dag}}
                                   & \num{1e6} & \num{0} & \num{0} & \num{0} & -- & -- & -- \\
                                   & \num{1e7} & \num{0} & \num{0} & \num{0} & -- & --& -- \\
                                   & \num{1e8} & \num{-4} & \num{1} & \num{-1.6939} & -- & -- & -- \\
                                   & \num{1e9} & \num{-19} & \num{44} & \num{1.2870e+01} & -- & -- & -- \\
        \hline
                                   & \num{1e6} & \num{-1} & \num{0} & \num{-4.9671e-01} & \num{2} & \num{2} & \num{2} \\
        Clock skew compensation with & \num{1e7} & \num{-1} & \num{0} & \num{-4.9793e-01} & \num{3} & \num{4} & \num{3.0049} \\
        approximate bounds of~(11) of~\cite{Kim:22-1}\tnote{\ddag} & \num{1e8} & \num{-1} & \num{1} & \num{-2.0906e-01} & \num{21} & \num{22} & \num{2.1005e+01} \\
                                   & \num{1e9} & \num{-1} & \num{1} & \num{-9.8559e-02} & \num{201} & \num{202} & \num{2.0101e+02} \\
        \hline
                                   & \num{1e6} & \num{-1} & \num{0} & \num{-4.9671e-01} & \num{1} & \num{2} & \num{1.5033} \\
        Clock skew compensation with & \num{1e7} & \num{-1} & \num{0} & \num{-4.9793e-01} & \num{1} & \num{8} & \num{4.5145} \\
        practical bounds of~(14) of~\cite{Kang:23} & \num{1e8} & \num{-1} & \num{1} & \num{-2.0906e-01} & \num{1} & \num{72} & \num{3.6647e+01} \\
                                   & \num{1e9} & \num{-1} & \num{1} & \num{-9.8559e-02} & \num{1} & \num{832} & \num{4.1968e+02} \\
        \hline

                                   & \num{1e6} & \num{0} & \num{0} & \num{0} & \num{1} & \num{1} & \num{1} \\
        Direct search & \num{1e7} & \num{0} & \num{0} & \num{0} & \num{1} & \num{1} & \num{1} \\
        based on Theorem~\ref{thm:csc-ds} & \num{1e8} & \num{0} & \num{0} & \num{0} & \num{1} & \num{4} & \num{2.4992} \\
                                   & \num{1e9} & \num{0} & \num{0} & \num{0} & \num{1} & \num{45} & \num{1.9132e+01} \\
        \hline
      \end{tabular}
    }%
    \begin{tablenotes}
    \item[*] With respect to ${\lfloor}\fp(i\frac{D}{A}){+}0.5{\rfloor}$ based
      on double precision; according to~\cite{Kang:23}, the results are the same
      as those based on much higher precision, including binary512 of IEEE
      754-2008.
    \item[\dag] ${\lfloor}\fp(i\frac{D}{A}){+}0.5{\rfloor}$ based on single
      precision.
    \item[\ddag] With $\varepsilon$ of~(11) of~\cite{Kim:22-1} set to
      $\num{1e-7}i$.
    \end{tablenotes}
  \end{threeparttable}
\end{table*}
We fix $D$ to $1,000,000$ and generate one million samples of $A$ corresponding
to clock skew uniformly distributed in the range of
$[{-}100\,\text{ppm},100\,\text{ppm}]$. The minimum and the maximum values of
$i$ (i.e., the hardware clock) in Table~\ref{tab:csc_results} correspond to
\SI{1}{\s} and \SI{1000}{\s}, respectively, at a sensor node whose clock
resolution is \SI{1}{\us}. For the clock skew compensation by single and
double-precision floating-point arithmetic, we round half up the results to
obtain integer values throughout the experiments.

As observed in \cite{Kang:23}, the clock skew compensation based on the extended
Bresenham's algorithm with approximate and practical bounds provides exactly the
same bounded compensation errors, while the numbers of iterations with the
practical bounds of~\cite{Kang:23} are larger than those with the approximate
bounds of~\cite{Kim:22-1} for $i{\geq}\num{1e7}$. On the other hand, the clock
skew compensation errors of the direct search algorithm are zero with much
smaller number of iterations for all the cases considered. Specifically, for
$i{\leq}\num{1e7}$ when the errors of the clock skew compensation by
single-precision floating-point arithmetic are zero, the direct search algorithm
does not incur unnecessary iterations unlike the clock skew compensation based
on the extended Bresenham's algorithm. Even when the initial value of the clock
compensated by single-precision floating-point arithmetic (i.e., $k$ in
\eqref{eq:csc-ds-k}) differs from that of double-precision floating-point
arithmetic, the proposed algorithm can minimize the number of iterations by
directly searching from the given initial value rather than jumping back in time
by the difference between the upper and lower bounds and starting with the
worst-case value as illustrated in Fig.~\ref{fig:csc-example}.

\section{Conclusions}
\label{sec:conclusions}
In this letter, we have proposed a new incremental error algorithm, called
direct search, for clock skew compensation immune to floating-point precision
loss. To address the computational complexity issues related with the use of the
bounds on the initial value of the clock in the clock skew compensation based on
the extended Bresenham's algorithm, the proposed algorithm starts from the given
initial value in searching for an optimally-skew-compensated clock and thereby
minimizes the number of iterations. Having theoretically established the direct
search algorithm, we have demonstrated through numerical examples that it can
significantly reduce the number of iterations in comparison to the prior work
based on the extended Bresenham's algorithm while completely eliminating the
effect of floating-point precision loss on clock skew compensation.

\appendix[Proof of Theorem~\ref{thm:csc-ds}]
\label{sec:proof-theorem-1}
\noindent
\textit{Cases 1, 2.1, and 3.1}: These are special cases of
Lemma~\ref{lem:scc-cond} where the rhs of \eqref{eq:scc-2} is zero.

Consider Case~2.1 for example. Because
\[
  \Delta - A = \left(kA - iD\right) -A = \left(k - 1\right)A - iD = 0,
\]
we have from Lemma~\ref{lem:scc-cond}
\[
  \left|jA-iD\right| \leq \left|(k-1)A-iD\right| = 0,
\]
which results in $j{=}k{-}1$. The proof of Cases~1 and 3.1 is similar.

\medskip
\noindent
\textit{Case 2.2}: It suffices to show that $|\td|$ corresponding to $k{-}1$
(i.e., the update) is smaller than that corresponding to $k$. Because
\[
  0 < \td-A = (k-1)A-i\frac{D}{A} < kA-i\frac{D}{A},
\]
we have
\[
  0 < |\td-A| = \left|(k-1)A-i\frac{D}{A}\right| < \left|kA-i\frac{D}{A}\right|.
\]

\medskip
\noindent
\textit{Case 2.3}: We cannot decrease $|\td|$ by increasing $k$, because
\[
  0 < \td = kA - i\frac{D}{A} < (k+1)A - i\frac{D}{A}.
\]
Likewise, we cannot decrease $|\td{-}A|$ by decreasing $k{-}1$, because
\[
  0 > \td - A = (k-1)A - i\frac{D}{A} > (k-2)A - i\frac{D}{A}.
\]
So, $j$ is either $k$ or $k{-}1$. If $|\td{-}A|<|\td|$, $j{=}k{-}1$ because
\[
  |\td-A| = |(k-1)A-iD| < |kA-iD|.
\]
Otherwise, $j{=}k$ because
\[
  |kA-iD| \leq |\td-A| = |(k-1)A-iD|.
\]

\medskip
\noindent
\textit{Case 3.2}: We cannot decrease $|\td|$ by decreasing $k$, because
\[
  0 > \td = kA-i\frac{D}{A} > (k-1)A-i\frac{D}{A}.
\]
Likewise, we cannot decrease $|\td{+}A|$ by increasing $k{+}1$, because
\[
  0 < \td+A = (k+1)A-i\frac{D}{A} < (k+2)A-i\frac{D}{A}.
\]
So, $j$ is either $k$ or $k{+}1$. If $|\td{+}A|<|\td|$, $j{=}k{+}1$ because
\[
  |\td+A| = |(k+1)A-iD| < |kA-iD|.
\]
Otherwise, $j{=}k$ because
\[
  |kA-iD| \leq |\td+A| = |(k+1)A-iD|.
\]

\medskip
\noindent
\textit{Case 3.3}: As in Case~2.2, it suffices to show that $|\td|$
corresponding to $k{+}1$ (i.e., the update) is smaller than that corresponding
to $k$. Because
\[
  0 > \td + A = (k+1)A - i\frac{D}{A} > kA - i\frac{D}{A},
\]
we have
\[
  0 < |\td+A| = \left|(k+1)A-i\frac{D}{A}\right| < \left|kA-i\frac{D}{A}\right|.
\]
\hfill\IEEEQED

\balance 


\end{document}